\documentclass[reprint,twocolumn,superscriptaddress,amsmath,amssymb,prl]{revtex4-1}
\usepackage{hyperref}

\usepackage{dcolumn}
\usepackage{epsfig}
\usepackage{graphics}
\usepackage[latin1]{inputenc} 
\usepackage[T1]{fontenc}
\usepackage{bm}
\usepackage{hyperref}
\newcommand{\ddst}{false}
\usepackage{color}
\usepackage{xcolor}
\usepackage{soul}

\begin{document}

\title{On the Allowable or Forbidden Nature of Vapor-Deposited Glasses}

 \author{Zhe Wang}
\affiliation{Physics of AmoRphous and Inorganic Solids Laboratory (PARISlab), Department of Civil and Environmental Engineering, University of California, Los Angeles, California 90095, USA}
 
 \author{Tao Du}
\affiliation{Physics of AmoRphous and Inorganic Solids Laboratory (PARISlab), Department of Civil and Environmental Engineering, University of California, Los Angeles, California 90095, USA}
 \affiliation{School of Civil Engineering, Harbin Institute of Technology, 150090 Harbin, China}
 
 \author{N. M. Anoop Krishnan}
 \affiliation{Department of Civil Engineering, Indian Institute of Technology Delhi, Hauz Khas, New Delhi 110016, India}
 \affiliation{Department of Materials Science and Engineering, Indian Institute of Technology Delhi, Hauz Khas, New Delhi 110016, India}
 
 \author{Morten M. Smedskjaer}
 \affiliation{Department of Chemistry and Bioscience, Aalborg University, 9220 Aalborg, Denmark}
 
   \author{Mathieu Bauchy}
\affiliation{Physics of AmoRphous and Inorganic Solids Laboratory (PARISlab), Department of Civil and Environmental Engineering, University of California, Los Angeles, California 90095, USA}
 \email[Contact: ]{bauchy@ucla.edu}
  
 
 \begin{abstract}
	Vapor deposition can yield glasses that are more stable than those obtained by the traditional melt-quenching route. However, it remains unclear whether vapor-deposited glasses are "allowable" or "forbidden," that is, if they are equivalent to glasses formed by cooling extremely slowly a liquid or if they differ in nature from melt-quenched glasses. Here, based on reactive molecular dynamics simulation (MD) of silica glasses, we demonstrate that the allowable or forbidden nature of vapor-deposited glasses depends on the temperature of the substrate and, in turn, is found to be encoded in their medium-range order structure.
\end{abstract}

\maketitle

If quenched fast enough, liquids can avoid crystallization and remain in the metastable supercooled liquid state \cite{Debenedetti2001Supercooled}. At the glass transition, the relaxation time eventually exceeds the observation time---so that melts experience a kinetic arrest and enter the out-of-equilibrium glassy state \cite{Varshneya1993Fundamentals, Zanotto2017glassy}. As out-of-equilibrium phases, the structure and properties of glasses depend on their history. In particular, the use of lower cooling rates results in the formation of more stable glasses that occupy lower states in the energy landscape \cite{Debenedetti2001Supercooled}. As an alternative route to melt-quenching, vapor deposition can yield ultrastable glasses \cite{Swallen2007Organic, Ediger2017Perspective:}---the degree of stability depending on the substrate temperature and deposition rate \cite{Swallen2007Organic, Kearns2007Influence, Singh2013Ultrastable,Kearns2008Hiking, Dalal2015Tunable}. The ultrastable nature of vapor-deposited glasses has been suggested to result from the enhanced mobility of the atoms at the surface of the deposited glass as compared to those in the bulk, thereby allowing deposited glasses to access lower energy states in an accelerated fashion \cite{Reid2016Age, Dutcher2008Glass}. However, it remains unclear whether ultrastable vapor-deposited glasses are $allowable$ (i.e., equivalent to glasses formed with a very slow cooling rate) or $forbidden$ (i.e., glasses that cannot be formed via any thermal route) \cite{Mauro2009Forbidden}. Although previous atomistic simulations have suggested that vapor-deposited and melt-quenched glasses exhibit a similar structure, they have thus far primarily been limited to model glasses, e.g., Lennard-Jones glasses \cite{Singh2013Ultrastable,Reid2016Age, Berthier2017Origin}.

Here, based on reactive MD simulations, we compare the structure of melt-quenched and vapor-deposited silica (SiO$_2$) glasses. Importantly, we demonstrate that vapor-deposited glasses are allowable in the case of high substrate temperatures, but forbidden for low substrate temperatures. We find that the forbidden nature of glasses deposited on low-temperature substrates is encoded in their ring size distribution.

To establish our conclusions, we conduct a series of MD simulations of vapor-deposited SiO$_2$ glasses. A tetragonal simulation box with a height of 75 \AA\ ($z$-axis) and lateral dimensions of 28 \AA\ ($x$- and $y$-axis) is first created. The box is surrounded by two reflective walls on top and bottom, while periodic boundary conditions are imposed laterally. A melt-quenched silica glass with a vertical thickness of 14 \AA\ is placed at the bottom and serves as substrate.  The deposition process is then simulated by iteratively placing new SiO$_2$ molecules at the top of the box (70 < $z$ < 75 \AA) with a downward velocity of 0.02 \AA/fs, wherein the initial horizontal position of inserted molecules is randomly chosen \cite{Singh2013Ultrastable, Reid2016Age}. We find that a deposition rate of 0.5 SiO$_2$/ps is slow enough to ensure a fair convergence of the potential energy of the deposited glass (see Supplementary Material). Hence, this deposition rate is kept constant in all simulations. The substrate temperature used herein ranges from 500 to 3500 K---as we observe that, at higher temperature, the inserted particles remain in a gas phase and do not deposit on the substrate. The dynamics of the deposited atoms is described in the microcanonical ensemble ($NVE$) coupled with a Langevin thermostat \cite{Schneider1978Molecular-dynamics}, as the use of the $NVE$ ensemble has been shown to yield more stable vapor-deposited configurations than those obtained within the canonical ($NVT$) ensemble \cite{Singh2013Ultrastable, Reid2016Age, Lyubimov2013Model}. The deposition process is continued until 512 SiO$_2$ molecules are deposited on the substrate, which results in the formation of a vapor-deposited glass that is about 35 \AA\ high. The vapor-deposited configuration is eventually subjected to an energy minimization using the FIRE algorithm \cite{Bitzek2006Structural} to obtain the inherent configuration \cite{Singh2013Ultrastable,Reid2016Age}. Note that, to avoid any spurious effect of the substrate and free surface, only a sub-portion of the vapor-deposited configuration is considered for all subsequent analysis (that is at least 5 \AA\ away from the substrate at the bottom and from the free surface at the top).

To compare vapor-deposited to melt-quenched SiO$_2$ glasses, we run some melt-quenching simulations \cite{Li2017Cooling}. First, initial configurations are created by randomly placing 512 SiO$_2$ molecules in a cube with periodic boundary conditions, while ensuring the absence of any unrealistic overlap. The system is then relaxed at 5000 K and zero pressure in the isothermal-isobaric ensemble ($NPT$) for 100 ps. The obtained liquid is then subsequently quenched into a glass by linearly cooling the system from 5000 to 300 K under zero pressure in the $NPT$ ensemble. Varying cooling rates ranging from 10,000 down to 0.1 K/ps are used to generate glasses exhibiting varying fictive temperatures \cite{Liu2018Glass}, i.e., differing thermal histories. All glasses are eventually subjected to an energy minimization to access their inherent configuration.

To ensure a consistent comparison between vapor-deposited and melt-quenched glasses, all simulations are conducted with the same forcefield. We adopt the reactive ReaxFF potential parameterized by Fogarty $et\ al$. \cite{Fogarty2010reactive} with a timestep of 0.5 fs. Importantly, ReaxFF can (i) account for charge transfers and dynamic formations of interatomic bonds \cite{Duin2001ReaxFF:}, (ii) handle coordination defects \cite{Duin2001ReaxFF:, Aktulga2012Parallel}, and (iii) realistically describe the structure of glassy SiO$_2$ \cite{Yu2016Revisiting}. Thanks to these features, ReaxFF can properly describe both the vapor deposition and melt-quenching processes with a constant set of parameters \cite{Meng2012Molecular}. All simulations are conducted with LAMMPS \cite{Plimpton1995Fast}.

\begin{figure}
	\begin{center}
		\includegraphics*[height=0.6\linewidth, keepaspectratio=true, draft=\ddst]{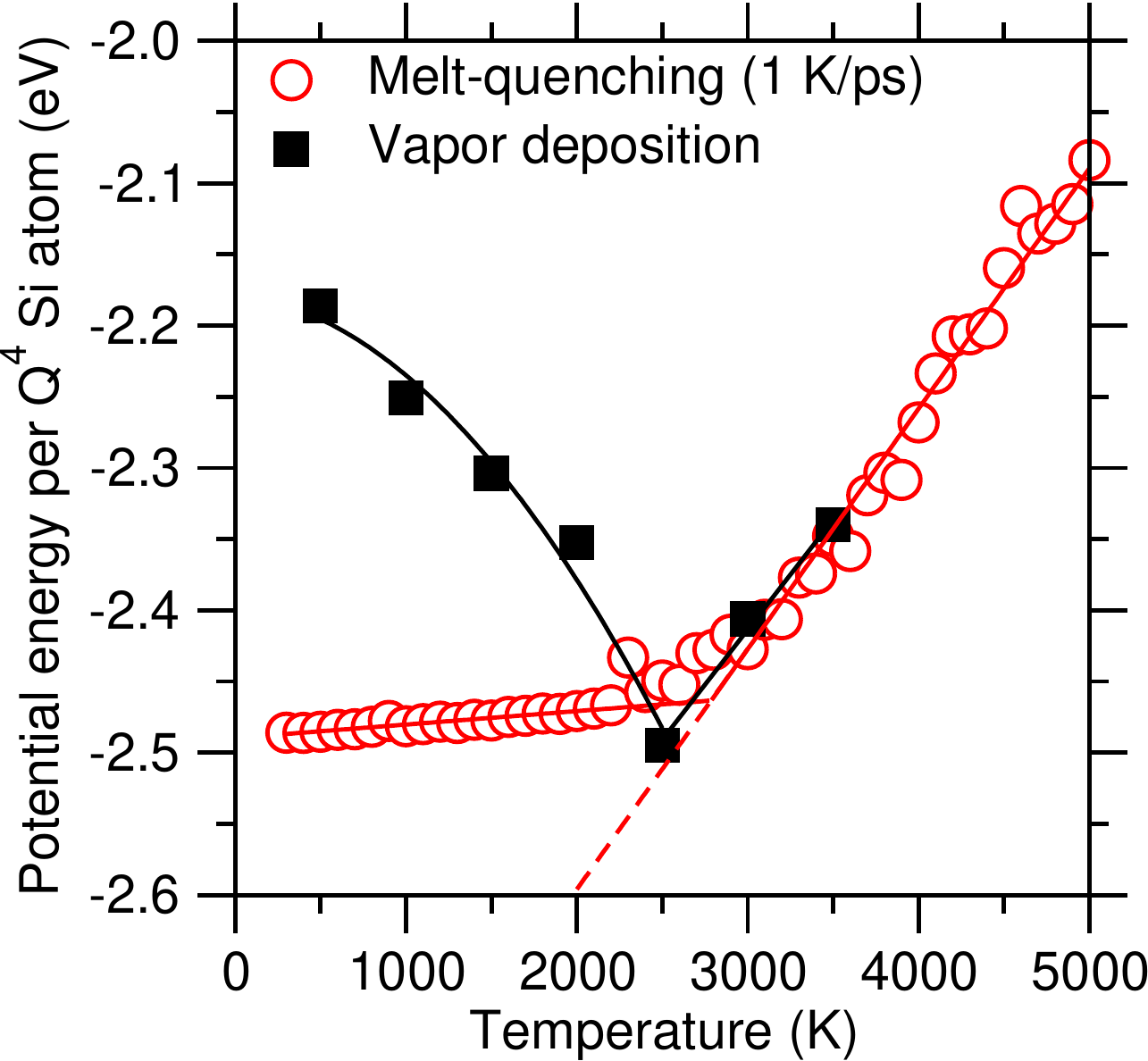}
		\caption{\label{fig:e4-T} Inherent structure average potential energy per Q$^4$ Si atom in both (i) vapor-deposited glasses as a function of the temperature of the substrate and (ii) a melt-quenched glass prepared with a cooling rate of 1 K/ps as a function of temperature. The solid lines are to guide the eye. The dashed line is an extrapolation of the supercooled liquid domain.
		}
	\end{center}
\end{figure}

We first assess the thermodynamic stability of the vapor-deposited glasses as a function of the substrate temperature. To this end, rather than relying on the total potential energy of the system, we compute the potential energy per Q$^4$ Si atom (i.e., Si atom connected to 4 bridging oxygen atoms) to filter out the contribution of coordination defects and isolate the intrinsic thermodynamic stability of the network. Figure 1 shows the inherent structure potential energy per Q$^4$ Si atom in vapor-deposited glasses as a function of the substrate temperature. The results are compared with the inherent structure potential energy of a melt-quenched glass prepared with a cooling rate of 1 K/ps \cite{Yu2016Revisiting, Krishnan2017Enthalpy, Krishnan2017Irradiation-driven}. Overall, we observe that the potential energy per Q$^4$ Si atom in vapor-deposited glasses exhibits a "V-shape" dependence on the substrate temperature, in agreement with previous results obtained for a 2D model glass \cite{Swallen2007Organic, Kearns2007Influence}. The total potential energy per atom exhibits a similar trend (see Supplementary Material). The most stable vapor-deposited glass is obtained for a substrate temperature of about 2500 K, which is slightly lower than the computed fictive temperature of the melt-quenched glass, that is, the temperature at which the energy exhibits a break in slope (see Fig. \ref{fig:e4-T})---note that both of these temperatures are here shifted toward higher values as compared to experiments due to the limited timescale accessible to MD simulations. This trend echoes previous simulation and experimental results \cite{Kearns2007Influence, Lyubimov2013Model}. Notably, at the substrate temperature of 2500 K, the vapor-deposited glass is slightly more stable than the melt-quenched glass---although this observation is specific to the deposition and cooling rates used herein. Overall, these results highlight that the behaviors of realistic (e.g., SiO$_2$) and model (e.g., Lennard Jones) vapor-deposited glasses appear to be governed by the same underlying physics.

\begin{figure}
	\begin{center}
		\includegraphics*[height=0.6\linewidth, keepaspectratio=true, draft=\ddst]{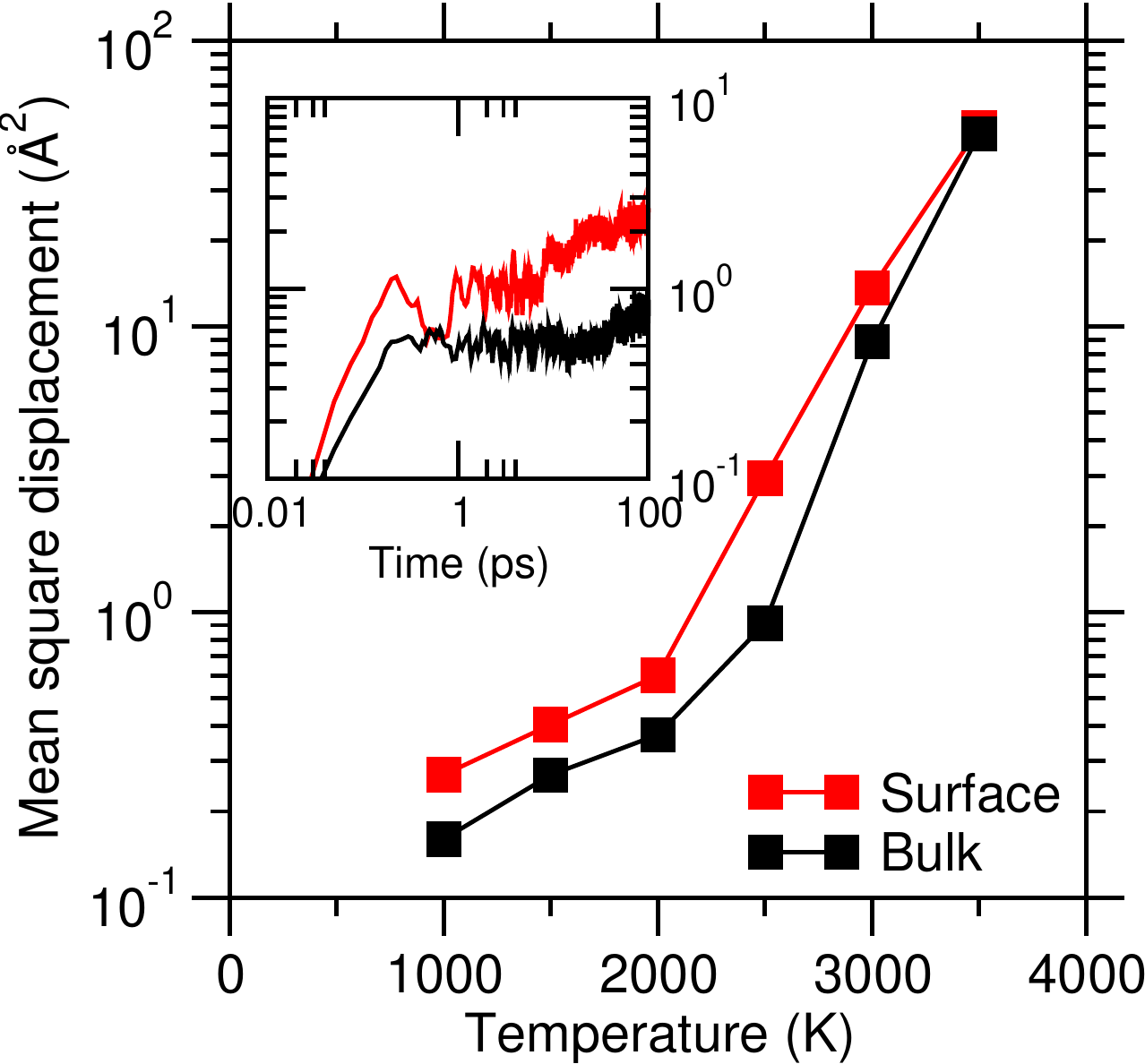}
		\caption{\label{fig:msd} Mean squared displacement (MSD) after 250 ps of dynamics of Si atoms located in the bulk or surface of vapor-deposited SiO$_2$ glasses as a function of temperature. The inset shows the MSD for bulk and surface Si atoms at 2500 K.
		}
	\end{center}
\end{figure}

We now investigate the origin of the high stability featured by vapor-deposited SiO$_2$ glasses at 2500 K (see Fig. \ref{fig:e4-T}). In line with results obtained for 2D model glasses \cite{Reid2016Age}, we suggest that the minimum of potential energy arises from a competition between thermodynamics and kinetics. To establish this picture, we explore the dynamics of the vapor-deposited glasses by computing the mean squared displacement (MSD) of the Si atoms as a function of temperature \cite{Lyubimov2013Model, Donati1999Spatial, Lin2014Molecular, Sun2017Structural}. All calculations are conducted in the $NVT$ ensemble over a duration of 250 ps. As expected, the MSD exhibits three stages, that is, (i) ballistic regime at short time (slope of 2 in log-log scale), (ii) cage-effect plateau at intermediate time, and (iii) diffusive regime, which manifests itself by a slope of 1 in log-log scale \cite{Bauchy2013Viscosity, Bauchy2011From} (see the inset of Fig. \ref{fig:msd} and Supplementary Material). Notably, the dynamics of the surface atoms (i.e., within a 5 \AA-thick region at the top of the sample) differs from those in the bulk. In details, we find that the MSD of the surface atoms is systematically larger than in the bulk (see Fig. \ref{fig:msd})---albeit to a lesser extent at higher temperature. We also observe that the duration of the cage effect is about one order of magnitude shorter at the surface than in the bulk (see the inset of Fig. \ref{fig:msd}). This likely arises from the fact that surface atoms are less constrained than bulk atoms \cite{Yu2018hydrophilic-to-hydrophobic} and, hence, have access to additional relaxation channels in the energy landscape \cite{Brian2013Surface}.

Based on these results, the V-shape of the potential energy can be rationalized as follows \cite{Reid2016Age}. At high temperature, the relaxation time of both vapor-deposited and melt-quenched systems is smaller than the observation time. Hence, both systems can reach the metastable equilibrium supercooled liquid state. As temperature decreases, lower-energy supercooled liquid states become more thermodynamically favored, so that both vapor-deposited and melt-quenched systems reach more stable positions in the energy landscape. However, at the vicinity of the glass transition, due to the kinetics slowdown, the relaxation time of bulk melt-quenched systems becomes longer than the observation time---so that they become out-of-equilibrium glasses and remain stuck in unstable positions in the energy landscape. In contrast, at constant observation time, vapor-deposited glasses relax faster than melt-quenched glasses thanks to the faster kinetics of their surface atoms. Hence, vapor-deposited glasses remain in the metastable equilibrium supercooled liquid state down to lower temperatures (and, hence, reach more stable basins in the energy landscape) than melt-quenched glasses at constant observation time. However, as the substrate temperature continues to decrease, the increased slowdown in relaxation kinetics eventually prevents the atoms from relaxing toward low-energy states when they get deposited at the glass surface, which results in an increase in potential energy. Overall, the substrate temperature at which vapor-deposited glasses feature minimum potential energy is controlled by the competition between thermodynamics (i.e., increased thermodynamic propensity to relax toward lower-energy states as temperature decreases) and kinetics (i.e., decreased ability to reach such stable states as temperature decreases).

\begin{figure}
	\begin{center}
		\includegraphics*[height=0.6\linewidth, keepaspectratio=true, draft=\ddst]{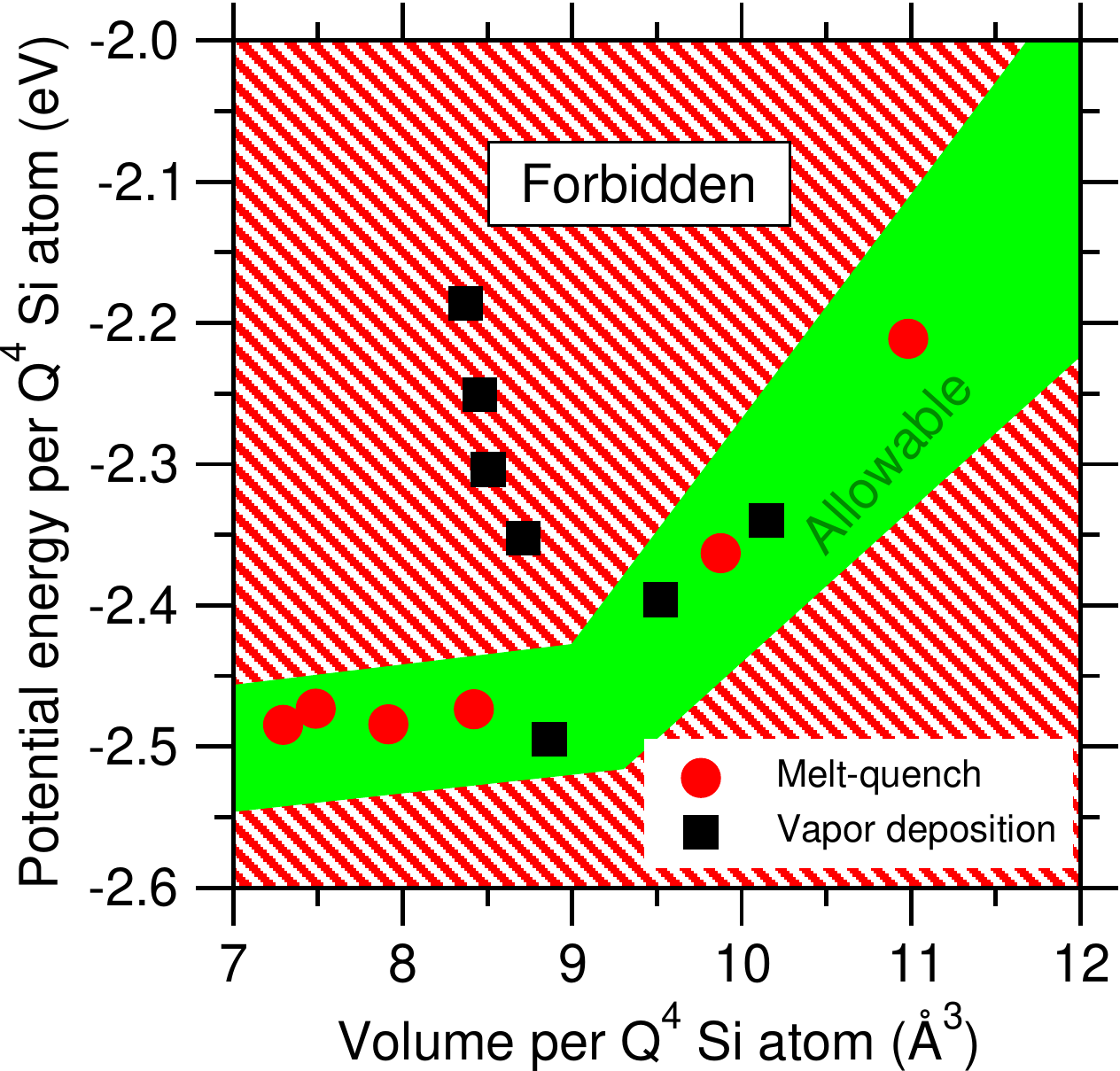}
		\caption{\label{fig:E-V} Inherent structure average potential energy as a function of the average Voronoi volume per Q$^4$ Si atom in (i) vapor-deposited glasses prepared with varying substrate temperature and (ii) melt-quenched glasses prepared with varying cooling rates (i.e., varying fictive temperature). The green region is a rough indication of the range of "allowable" states, whereas other states are "forbidden."
		}
	\end{center}
\end{figure}

We now interrogate whether vapor-deposited glasses are forbidden or allowable. That is, do vapor-deposited glasses differ in nature from melt-quenched glasses or can they also be formed by melt-quenching with a given (slow) cooling rate? Specifically, can the increase in the potential energy of vapor-deposited glasses with low substrate temperature (see Fig. \ref{fig:e4-T}) be understood as an increase in fictive temperature? To answer this question, Fig. \ref{fig:E-V} shows the inherent structure average potential energy as a function of the average Voronoi volume per Q$^4$ Si atom in vapor-deposited glasses prepared with varying substrate temperature and melt-quenched glasses prepared with cooling rates varying from 10,000 to 0.1 K/ps (i.e., varying fictive temperature) \cite{Krishnan2017Enthalpy}. Although the potential energy and volume do not uniquely characterize a glass \cite{Smedskjaer2015Unique}, the potential energy captures the degree of stability of a glass and, to the first order, largely depends on the short-range order, whereas the volume captures the overall compactness of the glass, which is strongly affected by the medium-range order \cite{Krishnan2017Enthalpy}. As such, the energy-volume space shown in Fig. \ref{fig:E-V} offers a convenient map to compare vapor-deposited and melt-quenched glasses.

We first focus on the melt-quenched glasses. For the range of cooling rates considered herein (which remain significantly larger than in typical experiments \cite{Li2017Cooling}), the average Voronoi volume per Q$^4$ Si atom decreases with decreasing cooling rate (i.e., the system becomes more optimally packed), while the average potential energy per Q$^4$ Si atom decreases and eventually plateaus (i.e., the system becomes more stable and achieves a lower fictive temperature) \cite{Krishnan2017Enthalpy}. These states define the range of allowable states that are accessible to melt-quenched glasses within the time scale accessible to our MD simulations (i.e., as roughly indicated by the green region in Fig. \ref{fig:E-V}).

We now place our attention to the states occupied by vapor-deposited glasses in the energy-volume map. We find that, at high substrate temperature, vapor-deposited glasses are equivalent to hyperquenched melt-quenched glasses prepared with high cooling rates (see Fig. \ref{fig:E-V}). This signals that, in this regime, vapor-deposited glasses are allowable and the increase in potential energy upon increasing substrate temperature can be understood in terms of an increase in fictive temperature. This echoes the fact that, in this range of temperature, both vapor-deposited and melt-quenched glasses are able to relax toward the same metastable equilibrium supercooled liquid state. In sharp contrast, at low substrate temperature, vapor-deposited glasses deviate from the states occupied by melt-quenched glasses in the energy-volume map (see Fig. \ref{fig:E-V}). Namely, upon decreasing substrate temperature, the potential energy per Q$^4$ Si atom increases while the volume per Q$^4$ Si atom keeps decreasing. This indicates that, in this regime, the increase in potential energy exhibited by vapor-deposited glasses upon increasing substrate temperature cannot be understood in terms of an increase in fictive temperature---so that such vapor-deposited glasses are forbidden. These results demonstrate that the allowable vs. forbidden nature of vapor-deposited glasses depends on the substrate temperature.

\begin{figure}
	\begin{center}
		\includegraphics*[height=0.6\linewidth, keepaspectratio=true, draft=\ddst]{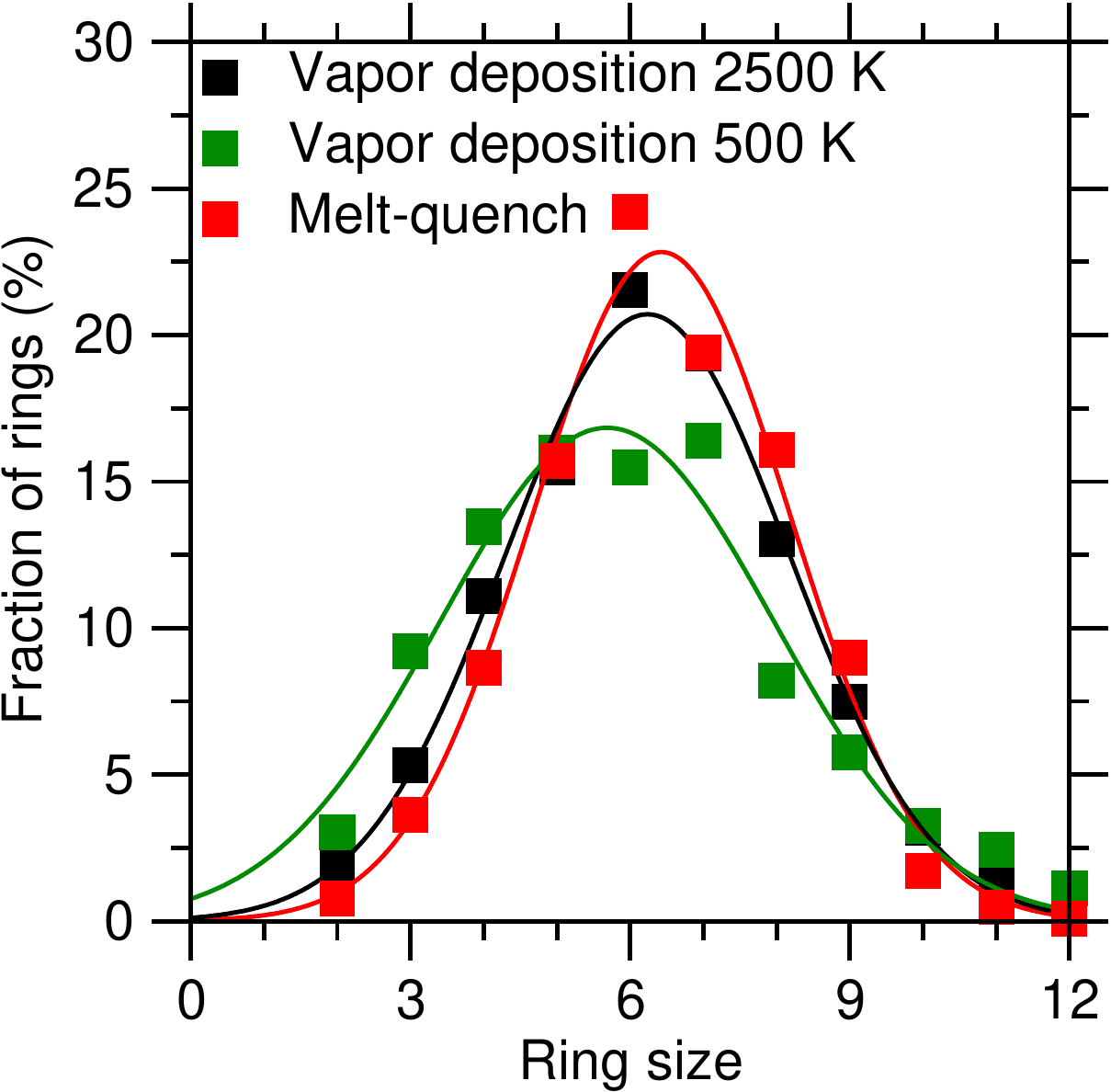}
		\caption{\label{fig:rings} Ring size distribution in vapor-deposited glasses prepared with a substrate temperature of 2500 K (allowable) and 500 K (forbidden). The ring size distribution in a melt-quenched glass prepared with a cooling rate of 1 K/ps is added for comparison. The lines are to guide the eye.
		}
	\end{center}
\end{figure}

Finally, we investigate how the allowable vs. forbidden nature of vapor-deposited glasses is encoded in their structure. First, we note that the pair distribution functions and bond angle distributions of vapor-deposited and melt-quenched glasses do not reveal any obvious differences (see Supplementary Material). This signals that the allowable vs. forbidden nature of vapor-deposited glasses is not encoded in their short-range order---which may explain why vapor-deposited glasses have previously been assumed to be structurally similar to melt-quenched ones \cite{Reid2016Age}. The short-range order analysis being inconclusive, we focus on the medium-range order, which, in silicate glasses, is described by the ring size distribution \cite{Micoulaut2003Rings, Geissberger1983Raman, le_roux_ring_2010, Elliott1991Medium-Range}---wherein a ring is defined as a closed path made of Si--O bonds in the network with a size being given by the number of Si atoms. All ring size distributions are computed using RINGS \cite{le_roux_ring_2010}.

Figure \ref{fig:rings} shows the ring size distribution of an allowable vapor-deposited glass prepared with a substrate temperature of 2500 K, i.e., the substrate temperature at which the deposited glass exhibits maximum stability. We find that, as expected, the distribution is centered around 5-to-6 membered rings \cite{Song2019Atomic}. No significant difference with the ring size distribution of a melt-quenched glass prepared with a cooling rate of 1 K/ps is observed (see Fig. \ref{fig:rings}). However, in contrast, we observe that the ring size distribution of forbidden vapor-deposited glasses (i.e., prepared with low substrate temperatures) exhibit distinct features. In details, we find that the ring size distribution of forbidden vapor-deposited glasses presents an excess of small rings (i.e., 4-membered rings and smaller) as compared to allowable glasses (see Fig. \ref{fig:rings}). Such small rings have been shown to be topologically overconstrained and to constitute a signature of instability \cite{Micoulaut2003Rings, Song2019Atomic}. In turn, such small rings result in the formation of efficiently-packed structures---since small rings are associated with low diameters, whereas larger rings present more open structures \cite{Shi2019Ring}. As such, the existence of a large fraction of small rings explains why forbidden vapor-deposited glasses prepared with low substrate temperatures simultaneously exhibit high potential energy and high packing efficiency. Such small rings can be formed when atoms get randomly deposited at the surface of the glass---irrespectively of the substrate temperature. However, due to their unstable nature, small rings are likely to quickly disappear as the surface atoms relax toward more stable configurations. However, the slowdown in relaxation kinetics experienced by vapor-deposited glasses prepared with low-temperature substrates prevents the efficient relaxation of such energetically-unfavorable small rings.

Overall, these results highlight that the forbidden or allowable nature of vapor-deposited glasses depends on the temperature of the substrate used during deposition and is controlled by a competition between thermodynamics and kinetics---wherein thermodynamics drives the relaxation of vapor-deposited glasses toward allowable metastable supercooled liquids, whereas kinetics can prevent such relaxation and tend to freeze some unrealistic small ring defects formed during deposition that are otherwise virtually absent from allowable melt-quenched glasses. More generally, these results suggest that the allowable vs. forbidden nature of disordered networks is encoded in their medium-range (rather than short-range) order. These results also suggest that, in addition to being a promising route toward the synthesis of ultrastable allowable glasses, vapor deposition offers an intriguing pathway toward the design of forbidden glasses that are not accessible to the melt-quench route and, hence, could exhibit unusual properties (e.g., enhanced mechanical properties, low propensity for relaxation, etc.).

\begin{acknowledgments}
This work was supported by the National Science Foundation under Grant No. 1928538. NMAK acknowledges some financial support provided by the Department of Science and Technology, India under the INSPIRE faculty scheme (DST/INSPIRE/04/2016/002774) and DST SERB Early Career Award (ECR/2018/002228).
\end{acknowledgments}


%

\end{document}